\documentclass[12pt]{iopart}
\pdfoutput=1
\usepackage{graphicx,color}
\begin{document}

\title{Occupation time statistics of the random acceleration model}

\author{Hermann Jo\"el Ouandji Boutcheng}
\address{Department of Physics, Faculty of Science, University of Yaounde I, P.O. Box 812, Cameroon; and The African Center of Excellence in Information and Communication Technologies (CETIC), Yaounde, Cameroon}
\author{Thomas Bouetou Bouetou}
\address{D\'epartement de Math\'ematiques et G\'enie Informatique, Ecole Nationale Sup\'erieure Polytechnique, P.O. Box 8390, Yaounde, Cameroon; and The African Center of Excellence 
in Information and Communication Technologies (CETIC), Yaounde, Cameroon}
\author{Theodore W. Burkhardt}
\address{Department of Physics, Temple University, Philadelphia, PA 19122, USA}
\author{Alberto Rosso}
\address{CNRS - Universit\'e Paris-Sud, LPTMS, UMR8626, 91405 Orsay Cedex, France}
\author{Andrea Zoia}
\address{Den-Service d'\'etudes des r\'eacteurs et de math\'ematiques appliqu\'ees (SERMA), CEA, Universit\'e Paris-Saclay, F-91191, Gif-sur-Yvette, France}
\ead{andrea.zoia@cea.fr}
\author{Kofane Timoleon Crepin}
\address{Department of Physics, Faculty of Science, University of Yaounde I, P.O. Box 812, Cameroon; and The African Center of Excellence in Information and Communication Technologies (CETIC), Yaounde, Cameroon}

\vspace{10pt}

\begin{indented}
\item[April  2016]
\end{indented}

\begin{abstract}
The random acceleration model is one of the simplest non-Markovian stochastic systems and has been widely studied in connection with applications in physics and mathematics. However, the occupation time and related properties are non-trivial and not yet completely understood. In this paper we consider the occupation time $T_+$ of the one-dimensional random acceleration model on the positive half-axis. We calculate the first two moments of $T_+$ analytically and also study the statistics of $T_+$ with Monte Carlo simulations. One goal of our work was to ascertain whether the occupation time $T_+$ and the time $T_m$ at which the maximum of the process is attained are statistically equivalent. For regular Brownian motion the distributions of $T_+$ and $T_m$ coincide and are given by L\'evy's arcsine law. We show that for randomly accelerated motion the distributions of $T_+$ and $T_m$ are quite similar but not identical. This conclusion follows from the exact results for the moments of the distributions and is also consistent with our Monte Carlo simulations.
\end{abstract}

%
\vspace{2pc}
%
\submitto{\JSTAT}
%
%
%

\section{Introduction}
\label{intro}

A variety of systems in physics, in the life and social sciences, and in engineering can be modeled in terms of particles traveling in a host medium which randomly change their state (position, direction, energy, etc.) in collisions with other particles or with the medium itself. The nature of the randomness may vary widely from one system to another. It may result either from the intrinsic stochastic nature of the underlying process or from uncertainty~\cite{duderstadt}. Some transport phenomena, while originating in deterministic and reversible events, can in practice only be described by resorting to the laws of probability.

A prominent example of such a stochastic system is the random acceleration model, which has been studied both in physics and mathematics. In physics it appears, for example, in the continuum description of the equilibrium statistics of a semiflexible polymer chain with non-zero bending energy~\cite{Burkhardt}. It also describes the steady state profile of a $(1+1)$-dimensional Gaussian interface~\cite{MB1} with dynamical exponent $z=4$ in the continuum version of the Golubovic-Bruinsma-Das Sarma-Tamborenea model~\cite{GBDT}. In addition, the random acceleration process arises in the description of the statistical properties of the Burgers equation with Brownian initial velocity~\cite{PV}.

The random  acceleration model is a non-trivial, non-Markov model, which is both relevant to real-world applications and simple enough so that it can be studied analytically. The first-passage properties and related properties have been investigated extensively over the last few decades~\cite{Burkhardt, MB1, McKean, Goldman, MW,  Sinai, Lachal, MB2, GL, Burkhardt2,Burkhardt5}. 
Recently, the extreme-value statistics of the process was analyzed, with special emphasis on the global maximum in a given time interval~\cite{Burkhardt,Gyorgy,Burkhardt3} and the time at which the global maximum is reached~\cite{zoia_rosso_majumdar_ram}. However, the occupation time statistics of random acceleration is still not understood in detail.

The occupation time in stochastic systems was first considered by mathematicians \cite{Levy,Kac,Lamperti,Cox} and has more recently been investigated in physical systems with continuous degrees of freedom and in connection with persistence (see \cite{review, Godreche} and references therein). In this paper we consider a randomly accelerated particle moving in one dimension on the infinite $x$ axis and study the occupation time $T_+$ on the positive $x$ axis. We calculate  the first two moments of $T_+$ analytically and also study the statistics of $T_+$ with Monte Carlo simulations. One of our aims was to learn whether the occupation time $T_+$ of the randomly accelerated particle and the time $T_m$ at which it attains its maximum displacement are statistically equivalent. Both our analytical and Monte Carlo results indicate that this is not the case. This is in contrast to regular Brownian motion, where the distributions of $T_+$ and $T_m$ coincide and are given by L\'evy's celebrated arcsine law~\cite{Levy,Feller}. 

The paper is structured as follows: In Sec.~\ref{sec_ram}, we derive partial differential equations which determine the moment generating function and the moments of the occupation time. In Sec.~\ref{sec_occupation_time} the first two moments of the occupation time $T_+$ are calculated explicitly and compared with the corresponding moments of the time $T_m$ at which the randomly accelerated particle makes its maximum excursion.  In Sec.~\ref{sec_numerical} we study the  moments of $T_+$ and its distribution with Monte Carlo simulations and compare the results with our analytic predictions for the first two moments of $T_+$ and  with exact results \cite{zoia_rosso_majumdar_ram} for the distribution of $T_m$. Sec.~\ref{sec_conclusions} contains concluding remarks. Some calculational details pertaining to Sec.~\ref{sec_ram} are given in~\ref{app_fk}.

\section{Differential equations for analyzing the occupation time}
\label{sec_ram}

The randomly accelerated particle we consider moves in one dimension according to the equations of motion
\begin{eqnarray}
& \frac{dx}{dt} = v,\label{eqmo1}\\
& \frac{dv}{dt} = \eta(t).\label{eqmo2}
\end{eqnarray}
Here $x(t)$ is the position of the particle, $v(t)$ is its velocity, and $\eta(t)$ is Gaussian white noise, with $\langle \eta(t) \rangle = 0$ and $\langle \eta(t) \eta(t') \rangle = 2\gamma \delta(t-t') $, $\gamma >0$. The initial conditions at time $t=0$ are $x(0) = x_0$ and $v(0) = v_0$.

The occupation time $T_A(t|x_0,v_0)$, i.e., the time the particle spends in a region $A$ during a total time of observation $t$, is formally expressed by the random variable
\begin{equation}
T_A(t|x_0,v_0) = \int_0^t V_A(x(t')) dt',\label{TAdef}
\end{equation}
where the marker function $V_A(x(t)) = 1$ if $x(t) \in A$ and vanishes otherwise. In studying the statistics of $T_A(t|x_0,v_0)$, it is convenient to introduce the moment generating function
\begin{equation}
Q_t(s|x_0,v_0) = \langle e^{-s T_A(t|x_0,v_0)} \rangle,
\label{gen_function}
\end{equation}
where $s$ is the variable conjugate to $T_A$ and has the dimensions of inverse time. The moments of the occupation time can be obtained by differentiating equation (\ref{gen_function}) with respect to $s$ and then setting $s=0$, according to
\begin{equation}
\langle T^n_A \rangle_t(x_0,v_0) = (-1)^n \frac{\partial^n}{\partial s^n} Q_t(s|x_0,v_0) \vert_{s=0}\thinspace.\label{TmfromQ}
\end{equation}

The evolution of $Q_t(s|x_0,v_0)$ is governed by a backward partial differential equation of the Fokker-Planck type, which is derived in~\ref{app_fk} and is given by
\begin{equation}
\frac{\partial}{\partial t}Q_{t} = v_0 \frac{\partial}{\partial x_0} Q_{t}+  \gamma \frac{\partial^2}{\partial v^2_0} Q_{t}- s V_A(x_0) Q_{t}.\label{diffeqQ}
\end{equation}
Taking derivatives of equation (\ref{diffeqQ}) with respect to $s$ and making use of equation (\ref{TmfromQ}) leads to the corresponding differential equation
\begin{equation}
\frac{\partial}{\partial t}\langle T^n_A \rangle_t= v_{0}\frac{\partial}{\partial x_{0}}\langle T^n_A \rangle_t  + \gamma \frac{\partial^{2}}{\partial v_{0}^{2}}\langle T^n_A \rangle_t + n V_A(x_{0})\langle T^{n-1}_A \rangle_t \label{diffeqTm}
\end{equation}
for the $n^{\rm th}$ moment of the occupation time. Note that the rightmost term or source term in equation (\ref{diffeqTm}) depends on the moment of order $n-1$.

With the initial condition 
$\langle T^n_A \rangle_0(x_0,v_0)= 0$, stemming from $T_A(t=0|x_0,v_0) = 0$, differential equation (\ref{diffeqTm}) has the explicit solution
\begin{eqnarray}
&&\langle T^n_A \rangle_t(x_0,v_0)\nonumber\\
&&\quad =n\int_{0}^{t} dt' \int _{-\infty}^\infty dx' \int_{-\infty}^\infty  dv'\thinspace V_A(x')\langle T^{n-1}_A \rangle_{t'} (x',v')
 G_{t-t'}(x',v';x_{0},v_{0}).\label{greenfunsol}
\end{eqnarray}
Here $G_{t}(x,v;x_{0},v_{0})$ is the Green's function satisfying
\begin{equation}
\frac{\partial}{\partial t} G_{t}(x,v;x_{0},v_{0}) = v_{0}\frac{\partial}{\partial x_{0}}G_{t}(x,v;x_{0},v_{0}) + \gamma \frac{\partial^{2}}{\partial v_{0}^{2}} G_{t}(x,v;x_{0},v_{0}),\label{diffeqG}
\end{equation}
with initial condition $G_{0}(x,v;x_{0},v_{0}) = \delta(x-x_0) \delta(v-v_0)$ and with boundary conditions that depend on the problem of interest.

The hierarchical relation (\ref{greenfunsol}) is the main result of this section that we will need below. It generates all the moments $\langle T^n_A \rangle_t(x_0,v_0)$ for positive integer $n$ recursively from the zeroth moment
\begin{equation}
\langle T^0_A \rangle_t(x_0,v_0) = Q_t(s|x_0,v_0) \vert_{s=0} = 1\label{T0fromQ}
\end{equation}
implied by equations (\ref{gen_function}) and (\ref{TmfromQ}).
We note that equation (\ref{greenfunsol}) also follows, without recourse to the differential equations ({\ref{diffeqQ}) and (\ref{diffeqTm}), from the definition (\ref{TAdef}) of $T_A$ and the interpretation of $G_{t}(x,v;x_{0},v_{0})$ as the probability density in the phase space $(x,v)$ for propagation from $(x_0,v_0)$ to $(x,v)$ in a time $t$.

\begin{figure}
\begin{center}
  \includegraphics[width=8cm]{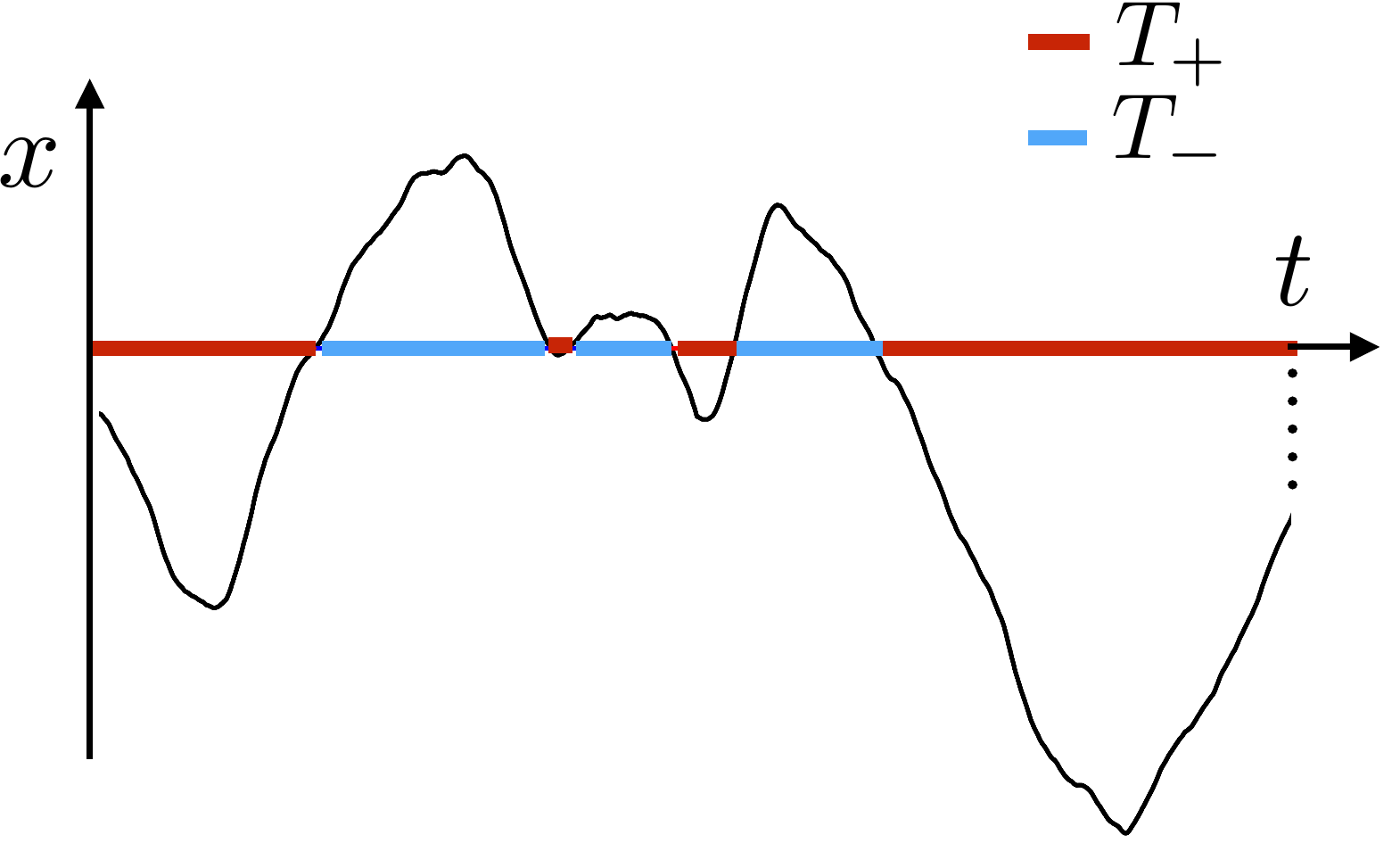}
\caption{Possible trajectory of a randomly acceleration particle moving on the $x$ axis with position $x_0$ and velocity $v_0$ at $t_0=0$. The occupation time $T_+(t|x_0,v_0)$ of the trajectory is the time spent by the particle on the positive half axis in a total time of observation $t$. Note that $T_-=t-T_+$.}
\label{fig1}
\end{center}
\end{figure}

\section{Occupation time on the half-line}
\label{sec_occupation_time}

For a particle which is free to move on the entire real axis, the Green's function in equations (\ref{greenfunsol}) and (\ref{diffeqG}) is given by~\cite{Burkhardt}
\begin{eqnarray}
&&G_{t}(x,v;x_0,v_0) = \frac{3^{1/2}}{2\pi \gamma t^{2}}\nonumber\\
&&\quad\times\exp\left\{-[3(x - x_0 - v t)(x - x_0 - v_0 t)/(\gamma t^3) + (v - v_0)^{2}/(\gamma t)]\right\}.\label{greenfunc}
\end{eqnarray}
Let us focus on the occupation time $T_+(t|x_0,v_0)$ that a randomly accelerated particle with initial position and velocity $x_0$ and $v_0$ spends on the positive half-axis $A=[0, +\infty)$ in a total time of observation $t$. The occupation time $T_+$ for a possible trajectory of the particle is illustrated in figure \ref{fig1}. Combining equations (\ref{greenfunsol}), (\ref{T0fromQ}), and (\ref{greenfunc}), we obtain
\begin{equation}
\langle T_+ \rangle_t (x_{0},v_{0})= \int_{0}^{t} dt' \int_0^\infty dx' \int_{-\infty}^{\infty} dv'\thinspace G_{t-t'}(x',v';x_{0},v_{0})\label{avT1p}
\end{equation}
for the average occupation time. The integrals over $x'$ and $v'$ can be evaluated explicitly, yielding
\begin{equation}
\langle T_+ \rangle_t(x_{0},v_{0})= 
 \frac{t}{2} + \frac{1}{2} \int_{0}^{t} dt'\thinspace \mbox{erf} \left[\frac{\sqrt{3}
}{2}\thinspace \frac{x_0+v_0(t- t')}{\sqrt{\gamma}\thinspace (t-t')^{3/2}} \right].
\end{equation}
In the special case $x_0 = 0$ and $v_0=0$, we obtain
\begin{equation}
\label{mean}
\langle T_+\rangle_t (x_{0}=0,v_{0}=0)=  \frac{1}{2}\thinspace t,\label{avT1final}
\label{eq:mean}
\end{equation}
as expected on physical grounds because of the symmetry of the process around the starting point.

According to equations (\ref{greenfunsol}) and (\ref{avT1p}), the second moment of $T_+$ is given by
\begin{eqnarray}
\langle T^2_+\rangle_t (x_{0},v_{0})=2 \int_{0}^{t} dt' \int_0^\infty dx' \int_{-\infty}^{\infty} dv' \int_{0}^{t'} dt'' \times \nonumber \\
\int_0^\infty dx'' \int_{-\infty}^{\infty} dv''\thinspace  G_{t'-t''}(x'',v'';x',v') G_{t-t'}(x',v';x_{0},v_{0}).
\end{eqnarray}
In the special case where $x_0 = 0$ and $v_0=0$, the integrals may be evaluated explicitly. First evaluating the Gaussian integrals over $v''$ and $v'$ and then integrating over $x''$ and $x'$, we obtain
\begin{eqnarray}
&&\langle T^2_+\rangle_t (x_{0}=0,v_{0}=0)= \frac{t^2}{4}\nonumber\\
&&\qquad\quad+ \frac{1}{\pi}\int_{0}^{t} dt' \int_{0}^{t'} dt'' \thinspace\tan^{-1}\left[\frac{ 2 t+t'-3 t''}{t'-t''}\;\sqrt{\frac{t-t'}{3 t+t'-4 t''}}\right].
\end{eqnarray}
The integrals over $t''$ and $t'$ can also be evaluated explicitly, and the final result is
\begin{equation}
\label{eq:main}
\langle T^2_+\rangle_t(x_{0}=0,v_{0}=0)= \frac{3^{3/2}}{4 \pi }\thinspace t^2\simeq 0.413497\thinspace t^2.\label{avT2final}
\label{eq:ratio}
\end{equation}

It is interesting to compare the exact results (\ref{avT1final}) and (\ref{avT2final}) for the first and second moments of the occupation time $T_+$ with the corresponding moments of the time $T_m$ at which the randomly accelerated particle makes its maximum excursion. As mentioned above, for regular Brownian motion the cumulative distributions of $T_+$ and $T_m$ coincide and are given by L\'evy's arcsine law~\cite{Levy}.

In reference~\cite{zoia_rosso_majumdar_ram}, the cumulative distribution of $T_m$ was derived analytically for the class of trajectories of a randomly accelerated particle which begin and end with velocity $v_i=v_f=0$. For this class of trajectories the random acceleration process corresponds to the integral of a Brownian bridge, and the cumulative distribution of the rescaled variable $z=T_m/t$ is given by
\begin{equation}
I_{{1\over 4},{1\over 4}}(z) = \frac{\Gamma({1\over 2})}{\Gamma({1\over 4})^2} B_z(\textstyle{{1\over 4},{1\over 4}}),\label{distTm}
\end{equation}
in terms of the incomplete beta function $B_z(p,q) = \int_0^z x^{p-1}(1-x)^{q-1}dx$. The $n^{\rm th}$ moment of $T_m$ for this cumulative distribution is
\begin{equation}
\langle T_m^n \rangle =t^n\int_0^1 dz\thinspace z^n {d\over dz}I_{{1\over 4},{1\over 4}}(z) = 
\frac{\Gamma({1\over 2})}{\Gamma({1\over 2}+n)} \frac{\Gamma({1\over 4}+n)}{\Gamma({1\over 4})}\;t^n,\label{Tmn}
\end{equation}
which implies
\begin{equation}
\langle T_m \rangle ={1\over 2}\thinspace t\thinspace,\qquad \langle T^2_m \rangle = {5\over 12}\thinspace t^2 \simeq 0.416667\thinspace t^2\label{Tm12}
\end{equation}
for the first and second moments.
Comparing equations (\ref{avT1final}), (\ref{avT2final}), and (\ref{Tm12}), we see that the first moments of $T_+$ and $T_m$ coincide and that the second moments differ, but by a small amount, less than $1\%$. Clearly, the cumulative distribution of $T_+/t$ is not given exactly by the expression in equation (\ref{distTm}), even though it appears to provide a very good approximation. The comparison between the distribution of the occupation time $T_+$ and the beta distribution has been considered by other researchers in the past: see, e.g.,~\cite{dornic, newman}.

\begin{figure}
\begin{center}
  \includegraphics[width=8cm]{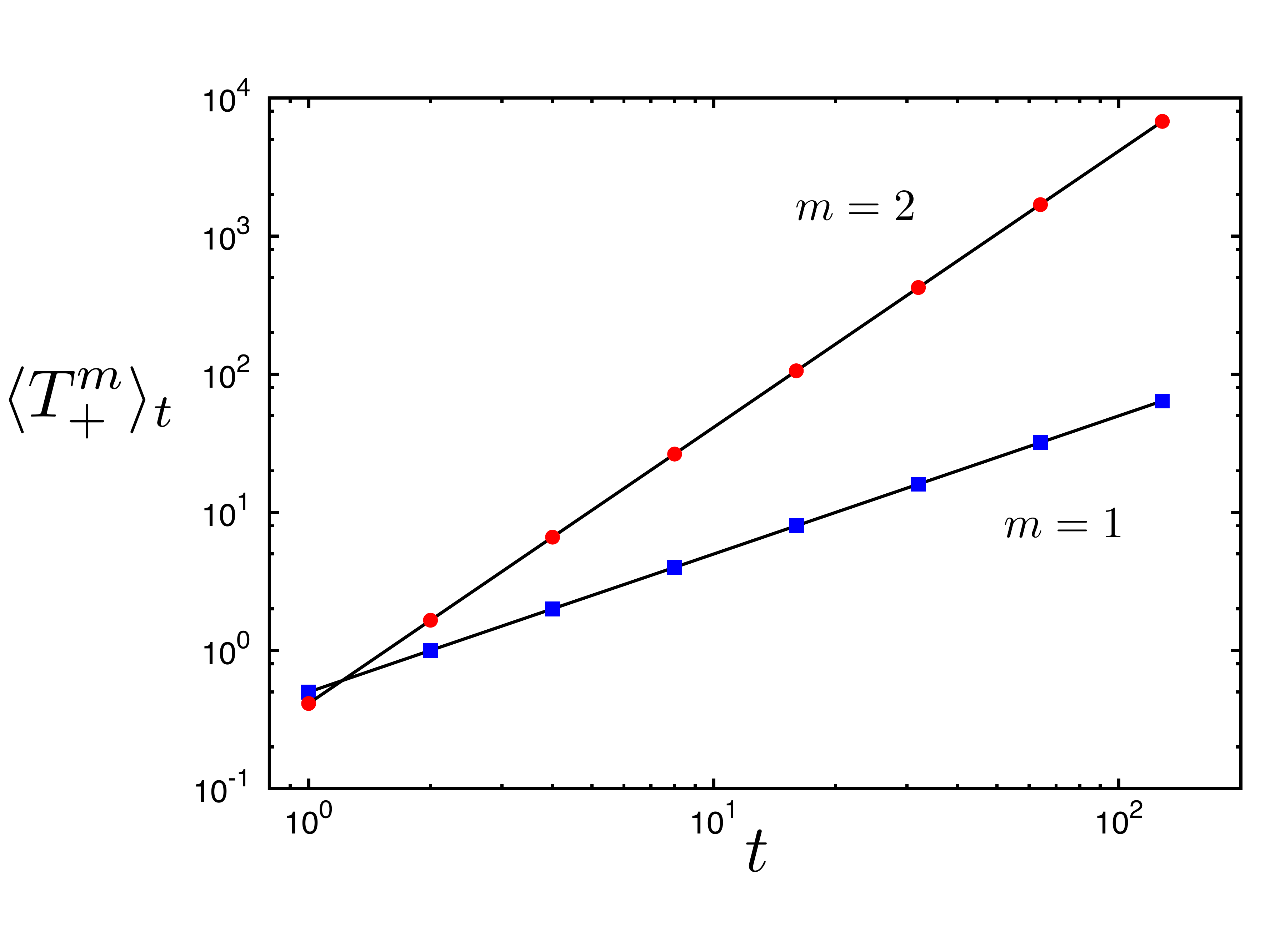}
\caption{First two moments of the occupation time $T_+$ for a randomly accelerated particle with initial conditions $x_0=0$ and $v_0=0$ 
as a function of the total time of observation $t$. The square and round points show our Monte Carlo results for the first and second moments, respectively, 
for $t= 2^0, \ldots,2^7$. Each point represents an average over $10^6$ realizations, and the error bars are smaller than the sizes of the points. The 
solid lines indicate the analytical predictions (\ref{eq:mean}) and (\ref{eq:main}) for the moments.}
\label{fig2}
\end{center}
\end{figure}

\section{Monte Carlo simulations}
\label{sec_numerical}

We have also studied the statistics of the occupation time $T_+$ of a randomly accelerated particle with Monte Carlo simulations. In the simulations the particle moves according to a discrete version of equations (\ref{eqmo1}) and (\ref{eqmo2}) given by
\begin{eqnarray}
& x_{t'+\Delta t} = x_{t'}+  v_{t'}\Delta t, \label{eqmo1discrete} \\
& v_{t'+\Delta t} =  v_{t'}+ \eta_{t'} \Delta t.\label{eqmo2discrete}
\end{eqnarray}
Here the $\eta_{t'}$ are independent and identically distributed (i.i.d.) Gaussian variables with zero mean and variance $2 \gamma\Delta t = 10^{-4}$. We set $\gamma=1$ in the simulations and chose the initial conditions $x_{t=0}=0$ and $v_{t=0}=0$ considered above. Our Monte Carlo results for the first two moments of the occupation time, based on $10^6$ realizations are compared with the exact analytical results in equations (\ref{avT1final}) and (\ref{avT2final}) in Fig.~\ref{fig2}. The agreement is excellent.

\begin{figure}
\begin{center}
  \includegraphics[width=8cm]{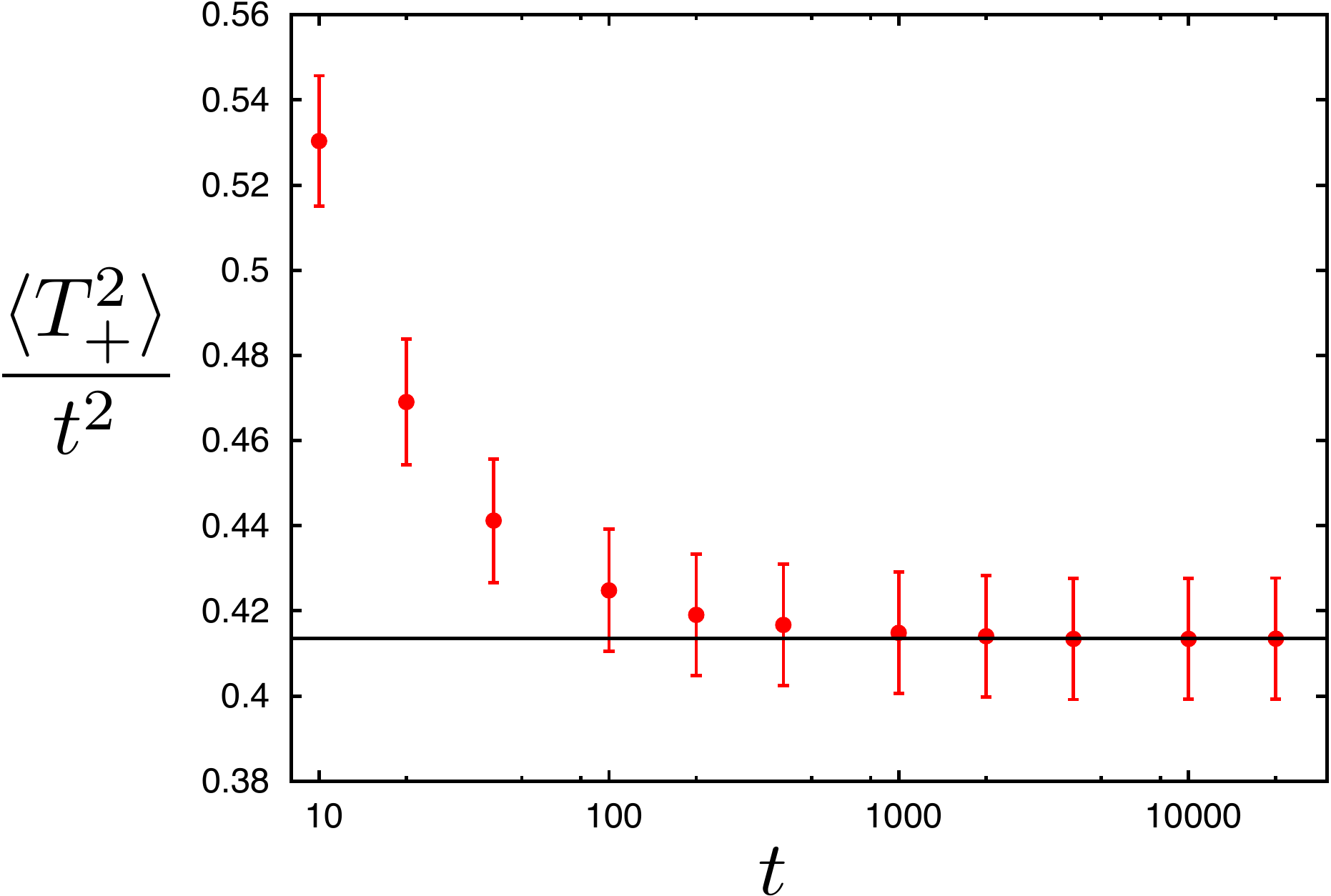}
\caption{Results for the pre-factor of the second moment $T^2_+$. We computed the ratio $T^2_+/t^2$ from Monte Carlo simulation, 
as described in Sec.~\ref{sec_numerical}. The points indicate the averages of $10^6$ realizations, and the solid line shows the prefactor 
$3^{3/2}/(4\pi)$ in equation (\ref{avT2final}).}
\label{fig3}
\end{center}
\end{figure}

We have also checked the prediction (\ref{avT2final}) for the second moment in another way. Setting $\Delta t=1$ in equations (\ref{eqmo1discrete}) and (\ref{eqmo2discrete}) and performing $10^6$ realizations, we computed the ratio $T^2_+/t^2$ for larger and larger values of $t$. As shown in Fig.~\ref{fig3}, the ratio saturates for large $t$ at a value indistinguishable from $3^{3/2}/(4\pi)$, in agreement with equation (\ref{avT2final}).

\begin{figure}
\begin{center}
  \includegraphics[width=8cm]{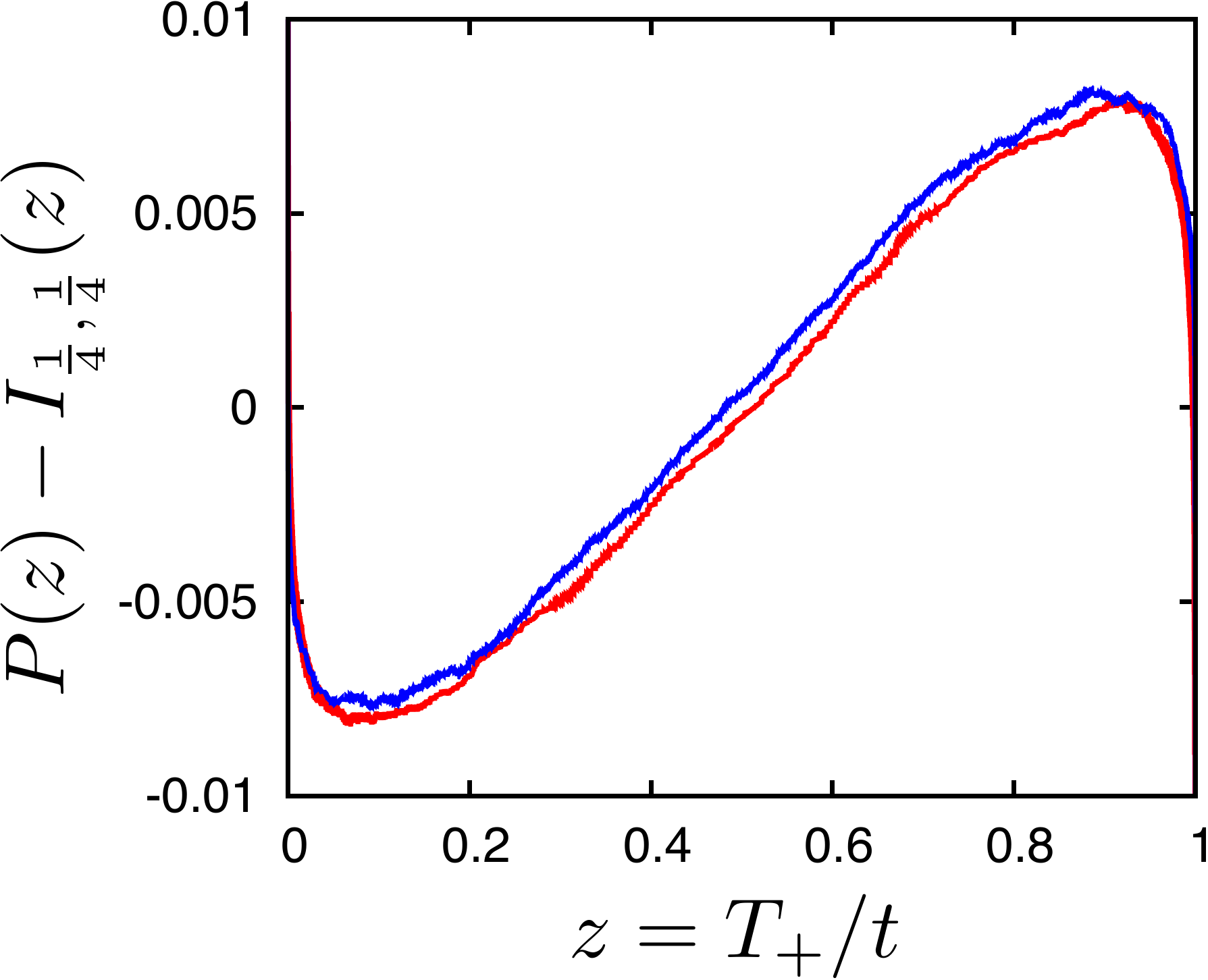}
\caption{Difference of the cumulative distribution $P(z)$ of the rescaled occupation time $z=T_+/t$, determined by Monte Carlo simulations, and the cumulative distribution $I_{{1\over 4},{1\over 4}}(z)$ given in equation (\ref{distTm}). On increasing the number of steps in the Monte Carlo simulations (blue: $10^4$ steps; red: $4\times 10^4$ steps), the difference converges to a non-trivial asymptotic shape and does not shrink to zero.}
\label{fig4}
\end{center}
\end{figure}

In addition to these studies of the first and second moments,  we have determined the complete cumulative distribution $P(z)$ of the rescaled occupation time $z=T_+/t$ numerically, from Monte Carlo simulations. The difference between the Monte Carlo result for $P(z)$  and the cumulative distribution $I_{{1\over 4},{1\over 4}}(z)$  in equation (\ref{distTm}) is plotted in Fig.~\ref{fig4}. From the figure it is clear that the two distributions are different but that the difference is small. This is the same conclusion we reached in Sec.~\ref{sec_occupation_time} on comparing exact results for the first and second moments of $T_+$ and $T_m$.

\section{Conclusions}
\label{sec_conclusions}

In this paper we have considered the occupation time statistics of a randomly accelerated particle moving in one dimension. After deriving evolution equations for the generating function and moments of the occupation time,  we calculated the first two moments of the occupation time $T_+$ on the positive half line exactly. Comparing these exact results with those for the first and second moments of the time $T_m$ at which the particle makes its maximum excursion, we conclude that the distributions of $T_+$ and $T_m$ are very similar but not identical, in contrast to the case of ordinary Brownian motion. Our Monte Carlo simulations of randomly accelerated motion are in excellent agreement with our analytical results for the first two moments of $T_+$ and confirm the conclusion that  $T_+$ and $T_m$ have very similar but not identical distributions. The calculation of higher moments of $T_+$ and its exact distribution are challenging problems for future study.


\appendix

\section{Derivation of differential equation (\ref{diffeqQ}) for the generating function}
\label{app_fk}

Considering a particle with position $x_0$ and velocity $v_0$ at time $t=0$ and following standard steps in deriving differential equations satisfied by path integrals (see, for example, \cite{Burkhardt,Kac,risken}), we decompose the total observation time $[0,t+dt]$ in a first interval from $t=0$ to $dt$ and a second interval from $dt$ to $t+dt$. Since the random acceleration process is Markovian in the two-dimensional phase space $(x,v)$, equations (\ref{eqmo1}) and (\ref{eqmo2}) imply
\begin{eqnarray}
Q_{t+dt}(s|x_0,v_0) = \langle e^{-s T_A(t+dt|x_0,v_0)} \rangle \nonumber \\
= \langle e^{-s \int_0^{dt} V_A(x(t')) dt'} e^{-s\int_{dt}^{t+dt} V_A(x(t')) dt'} \rangle.\label{Qdefappendix}
\end{eqnarray}
For infinitesimal $dt$ the quantity
\begin{equation}
e^{-s\int_0^{dt} V_A(x(t')) dt'} \to  e^{- s V_A(x_0) dt}
\end{equation}
in equation (\ref{Qdefappendix}) is completely deterministic and can be placed outside the angular brackets, yielding
\begin{equation}
Q_{t+dt}(s|x_0,v_0) = e^{- sV_A(x_0) dt} \langle e^{-s\int_{dt}^{t+dt} V_A(x(t')) dt'} \rangle.
\end{equation}
Translating the time in the integral on the right back by $dt$, we obtain
\begin{eqnarray}
Q_{t+dt}(s|x_0,v_0) = e^{- sV_A(x_0) dt} \langle e^{-s\int_{0}^{t} V_A(x(t')) dt'} \rangle\nonumber \\
 = e^{-s V_A(x_0) dt} \langle Q_{t}(s|x_0+\Delta x, v_0 + \Delta v) \rangle,
\label{eq_bracket}
\end{eqnarray}
where $\Delta x$ and $\Delta v$ are the changes in position and velocity in the time interval from $t=0$ to $dt$. On expanding the right-hand side for small $\Delta x$, $\Delta v$, and $dt$, equation (\ref{eq_bracket}) takes the form
\begin{eqnarray}
Q_{t+dt}(s|x_0, v_0)&-&Q_{t}(s|x_0, v_0 )=\left[-sV_A(x_0) + \Delta x  \frac{\partial}{\partial x_0}\right. \nonumber \\
&&\quad+\left. \Delta v  \frac{\partial}{\partial v_0} +\frac{1}{2} (\Delta v)^2  \frac{\partial^2}{\partial v^2_0}  + \cdots\right]Q_{t}(s|x_0, v_0 ). \label{expansion}
\end{eqnarray}
Substituting $\langle \Delta x  \rangle = v_0 dt$, $\langle \Delta v  \rangle = 0$ and $\langle (\Delta v)^2  \rangle = \gamma dt$ into equation (\ref{expansion}) and dividing the equation by $dt$ leads to the partial differential equation (\ref{diffeqQ}) for the evolution of the generating function. Since the derivatives on the right-hand side of equation (\ref{diffeqQ}) act on the initial coordinates $x_0$ and $v_0$, it is an example of a ``backward'' evolution equation.
\vskip 0.7cm

\noindent{\bf References}\vskip 0.2cm


\begin{thebibliography}{}
%
\bibitem{duderstadt} Duderstadt J J and Martin W R 1979 {\em Transport Theory} (New York: Wiley)

\bibitem{Burkhardt} Burkhardt T W 1993 Semiflexible Polymer in the Half Plane and Statistics of the Integral of a Brownian Curve {\em J. Phys. A: Math. Gen.} {\bf 26} L1157

\bibitem{MB1} Majumdar S N and Bray A J 2001 Spatial Persistence of Fluctuating Interfaces {\em Phys. Rev. Lett.} {\bf 86} 3700

\bibitem{GBDT} Golubovic L and  Bruinsma R 1991 Surface diffusion and fluctuations of growing interfaces. {\em Phys. Rev. Lett.}  {\bf 66} 321; Das Sarma S and Tamborenea P 1991 A new universality class for kinetic growth: One-dimensional molecular-beam epitaxy {\em Phys. Rev. Lett.} {\bf 66} 325 

\bibitem{PV} Valageas P 2009 Statistical properties of the Burgers equation with Brownian initial velocity {\em J. Stat. Phys.} {\bf 134} 589 

\bibitem{McKean} McKean H P 1963 A winding problem for a resonator driven by a white noise {\em J. Math. Kyoto Univ.} {\bf 2} 227

\bibitem{Goldman} Goldman M 1971 On the first passage of the integrated Wiener process {\em  Ann. Math. Stat.} {\bf 42} 2150

\bibitem{MW} Marshall T W and Watson E J 1985 A drop of ink falls from my pen...It comes to earth, I know not when {\em J. Phys. A: Math. Gen.} {\bf 18} 3531

\bibitem{Sinai} Sinai Y G 1992 Distribution of some functionals of the integral of a random walk {\em Theor. Math. Phys.} {\bf 90} 219

\bibitem{Lachal} Lachal A 1994 Last passage time for integrated Brownian motion {\em Stoch. Proc. Appl.} {\bf 49} 57

\bibitem{MB2} Majumdar S N and Bray A J 1998 Persistence with Partial Survival {\em Phys. Rev. Lett.} {\bf 81} 2626 

\bibitem{GL} De Smedt G, Godr\`eche C and Luck J M 2001 Partial survival and inelastic collapse for a randomly accelerated particle {\em Europhys. Lett.} {\bf 53} 438 

\bibitem{Burkhardt2} Burkhardt T W 2000 Dynamics of Absorption of a Randomly Accelerated Particle {\em J. Phys. A: Math. Gen.} {\bf 33} L429

\bibitem{Burkhardt5} Burkhardt T W 2014 First Passage of a Randomly Accelerated Particle {\em First-Passage Phenomena and Their Applications} ed R Metzler, G Oshanin and S Redner (Singapore: World Scientific)

\bibitem{Gyorgy} Gy\"{o}rgyi G, Moloney, N R, Ozog\'any K and R\'acz Z 2007 Maximal height statistics for 1/f$^{\alpha}$ signals {\em Phys. Rev. E} {\bf 75} 021123 

\bibitem{Burkhardt3} Burkhardt T W 2008 First-Passage and Extreme-Value Statistics of a Particle Subject to a Constant Force Plus a Random Force {\em J. Stat. Phys.} {\bf 133} 217

\bibitem{zoia_rosso_majumdar_ram} Majumdar S N, Rosso A and Zoia A 2010 Time at which the maximum of a random acceleration process is reached {\em J. Phys. A: Math. Theor.} {\bf 43} 115001

\bibitem{Levy} L\'evy P 1939 Sur certains processus stochastiques homog\`enes {\em Comp. Math.} {\bf 7} 283

\bibitem{Kac} Kac M 1949 On distributions of certain Wiener functionals {\em Trans. Am. Math. Soc.} {\bf 65} 1

\bibitem{Lamperti} Lamperti J 1958 An occupation time theorem for a class of stochastic processes {\em Trans. Am. Math. Soc.} {\bf 83} 380

\bibitem{Cox} Cox J T and Griffeath D 1985 Large deviations for some infinite particle system occupation times {\em Contemp. Math.} {\bf 41} 43

\bibitem{review} Bray A J, Majumdar, S N and Schehr G 2013 Persistence and first-passage properties in nonequilibrium systems {\em Advances in Physics} {\bf 62} 225

\bibitem{Godreche} Godr\`eche C and Luck J M 2001 Statistics of the occupation time for a random walk in the presence of a moving boundary 
{\em J. Phys. A: Math. Gen.} {\bf 34} 7153

\bibitem{Feller} Feller W 1970 {\em An Introduction to Probability Theory and its Applications} (New York: Wiley)

\bibitem{risken} Risken H 1989 {\em The Fokker-Planck Equation: Methods of Solution and Applications} 2nd ed (Berlin: Springer)

\bibitem{dornic} Godr\`eche C and Dornic I 1998 Large deviations and nontrivial exponents in coarsening systems {\em J. Phys. A} {\bf 31} 5413

\bibitem{newman} Newman T J and Toroczkai Z 1998 Diffusive persistence and the `sign-time' distribution {\em Phys. Rev. E} {\bf 58} R2685(R)

\end{thebibliography}
\end{document}